\documentclass[sigconf]{acmart}

\usepackage{booktabs} 

\setcopyright{rightsretained}

\copyrightyear{2017} 
\acmYear{2017} 
\setcopyright{acmcopyright}
\acmConference{GECCO '17}{July 15-19, 2017}{Berlin,
  Germany}
\acmPrice{15.00}
\acmDOI{http://dx.doi.org/10.1145/3071178.3071312}
\acmISBN{978-1-4503-4920-8/17/07}

\begin{document}
\title{Conversion Rate Optimization\\through Evolutionary Computation}

\author{Risto Miikkulainen$^{1,2}$, Neil Iscoe$^{1}$, Aaron
  Shagrin$^{1}$, Ron Cordell$^{1}$, Sam Nazari$^{1}$, Cory
  Schoolland$^{1}$, Myles Brundage$^{1}$, Jonathan Epstein$^{1}$,
  Randy Dean$^{1}$, Gurmeet Lamba$^{1}$
}
\affiliation{%
  \institution{$^1$Sentient Technologies, Inc.\\
    $^2$The University of Texas at Austin\\}
}

\begin{abstract}
  Conversion optimization means designing a web interface so that as
  many users as possible take a desired action on it, such as register
  or purchase. Such design is usually done by hand, testing one change
  at a time through A/B testing, or a limited number of combinations
  through multivariate testing, making it possible to evaluate only a
  small fraction of designs in a vast design space. This paper
  describes Sentient Ascend, an automatic conversion optimization
  system that uses evolutionary optimization to create effective web
  interface designs.  Ascend makes it possible to discover and utilize
  interactions between the design elements that are difficult to
  identify otherwise.  Moreover, evaluation of design candidates is
  done in parallel online, i.e.\ with a large number of real users
  interacting with the system.  A case study on an existing media site
  shows that significant improvements (i.e.\ over 43\%) are possible
  beyond human design. Ascend can therefore be seen as an approach to
  massively multivariate conversion optimization, based on a massively
  parallel interactive evolution.
\end{abstract}

%
%




\maketitle

\section{Introduction}
\label{sec:intro}

In e-commerce, designing web interfaces (i.e.\ web pages and
interactions) that convert as many users as possible from casual
browsers to paying customers is an important goal
\cite{ash:book12,salehd:book11}. While there are some well-known
design principles, including simplicity and consistency, there are
often also unexpected interactions between elements of the page that
determine how well it converts. The same element, such as a headline,
image, or testimonial, may work well in one context but not in
others---it is often hard to predict the result, and even harder to
decide how to improve a given page.

An entire subfield of information technology has emerged in this area,
called conversion rate optimization, or conversion science. The
standard method is A/B testing, i.e.\ designing two different versions
of the same page, showing them to different users, and collecting
statistics on how well they each convert \cite{kohavi:encyclopedia16}.
This process allows incorporating human knowledge about the domain and
conversion optimization into the design, and then testing their
effect. After observing the results, new designs can be compared and
gradually improved. The A/B testing process is difficult and
time-consuming: Only a very small fraction of page designs can be
tested in this way, and subtle interactions in the design are likely
to go unnoticed and unutilized. An alternative to A/B is multivariate
testing, where all value combinations of a few elements are tested at
once.  While this process captures interactions between these
elements, only a very small number of elements is usually included
(e.g.\ 2-3); the rest of the design space remains unexplored.

This paper describes a new technology for conversion optimization
based on evolutionary computation. This technology is implemented in
Ascend, a conversion optimization product by Sentient Technologies,
deployed in numerous e-commerce websites of paying customers since
September 2016 \cite{sentient:ascend17}. Ascend uses a
customer-designed search space as a starting point.  It consists of a
list of elements on the web page that can be changed, and their
possible alternative values, such as a header text, font, and color,
background image, testimonial text, and content order. Ascend then
automatically generates web-page candidates to be tested, and improves
those candidates through evolutionary optimization.

Because e-commerce sites often have high volume of traffic, fitness
evaluations can be done live with a large number of real users in
parallel. The evolutionary process in Ascend can thus be seen as a
massively parallel version of interactive evolution, making it
possible to optimize web designs in a few weeks. From the application
point of view, Ascend is a novel method for massively multivariate
optimization of web-page designs.  Depending on the application,
improvements of 20-200\% over human design have been observed through
this approach \cite{sentient:ascend17}.

This paper describes the technology underlying Ascend, presents
an example use case, and outlines future opportunities for
evolutionary computation in optimizing e-commerce.

\section{Background}
\label{sec:bg}

With the explosive growth of e-commerce in recent years, entirely new
areas of study have emerged. One of the main ones is conversion rate
optimization, i.e.\ the study of how web interfaces should be designed
so that they are as effective as possible in converting users from
casual browsers to actual customers. Conversion means taking a desired
action on the web interface such as making a purchase, registering for
a marketing list, or clicking on other desired link in an email,
website, or desktop, mobile, or social media application
\cite{ash:book12,salehd:book11}. Conversions are usually measured in
number of clicks, but also in metrics such as resulting revenue or
time spent on the site and rate of return to the site.

Conversions are currently optimized in a labor-intensive manual
process that requires significant expertise. The web design expert or
marketer first creates designs that s/he believes to be effective.
These designs are then tested in an A/B testing process, by directing
user traffic to them, and measuring how well they convert. If the
conversion rates are statistically significantly different, the better
design is adopted.  This design can then be improved further, using
domain expertise to change it, in another few rounds of creation and
testing.

Conversion optimization is a fast-emerging component of e-commerce. In
2016, companies spent over \$72 billion to drive customers to their
websites \cite{emarketer:sep16}. Much of that investment does not
result in sales: conversion rates are typically 2-4\% (i.e.\ 2-4\% of
the users that come to the site convert within 30 days). In 2014, only
18\% of the top 10,000 e-commerce sites did any conversion
optimization; in January 2017, 30\% of them did so
\cite{builtwith:jan17}. The growth is largely due to available
conversion optimization tools, such as Optimizely, Visual Website
Optimizer, Mixpanel, and Adobe Target \cite{builtwith:jan17}.  These
tools make it possible to configure the designs easily, allocate users
to them, record the results, and measure significance.

This process has several limitations. First, while the tools make the
task of designing effective web interfaces easier, the design is still
done by human experts. The tools thus provide support for confirming
the experts' ideas, not helping them explore and discover novel
designs. Second, since each step in the process requires statistical
significance, only a few designs can be tested. Third, each
improvement step amounts to one step in hillclimbing; such a process
can get stuck in local maxima.  Fourth, the process is aimed at
reducing false positives and therefore increases false negatives,
i.e.\ designs with good ideas may be overlooked.  Fifth, while the
tools provide support for multivariate testing, in practice only a few
combinations can be tested (e.g.\ five possible values for two
elements, or three possible values for three elements).  As a result,
it is difficult to discover and utilize interactions between
design elements.

Evolutionary optimization is well suited to address these limitations.
Evolution is an efficient method for exploration; only weak
statistical evidence is needed for progress; its stochastic nature
avoids getting stuck in local maxima; good ideas will gradually become
more prevalent. Most importantly, evolution searches for effective
interactions.  For instance, Ascend may find that the button needs to
be green, but *only* when it is transparent, *and* the header is in
small font, *and* the header text is aligned. Such interactions are
very difficult to find using A/B testing, requiring human insight into
the results.  Evolution makes this discovery process automatic. With
Ascend, it is thus possible to optimize conversions better and at a
larger scale than before.

Technically, Ascend is related to approaches to interactive evolution
\cite{takagi:ieee01,secretan:ec11} and crowdsourcing
\cite{brabham:book13,lehman:tpnc13} in that evaluations of candidates
are done online by human users. The usual interactive evolution
paradigm, however, employs a relatively small number of human
evaluators, and their task is to select good candidates or evaluate
the fitness of a pool of candidates explicitly. In contrast in Ascend,
a massive number of human users are interacting with the
candidates, and fitness is derived from their actions (i.e. convert or
not) implicitly.

\begin{figure*}[!t]
  \begin{minipage}{\columnwidth}
    \begin{center}
      \includegraphics[width=\columnwidth]{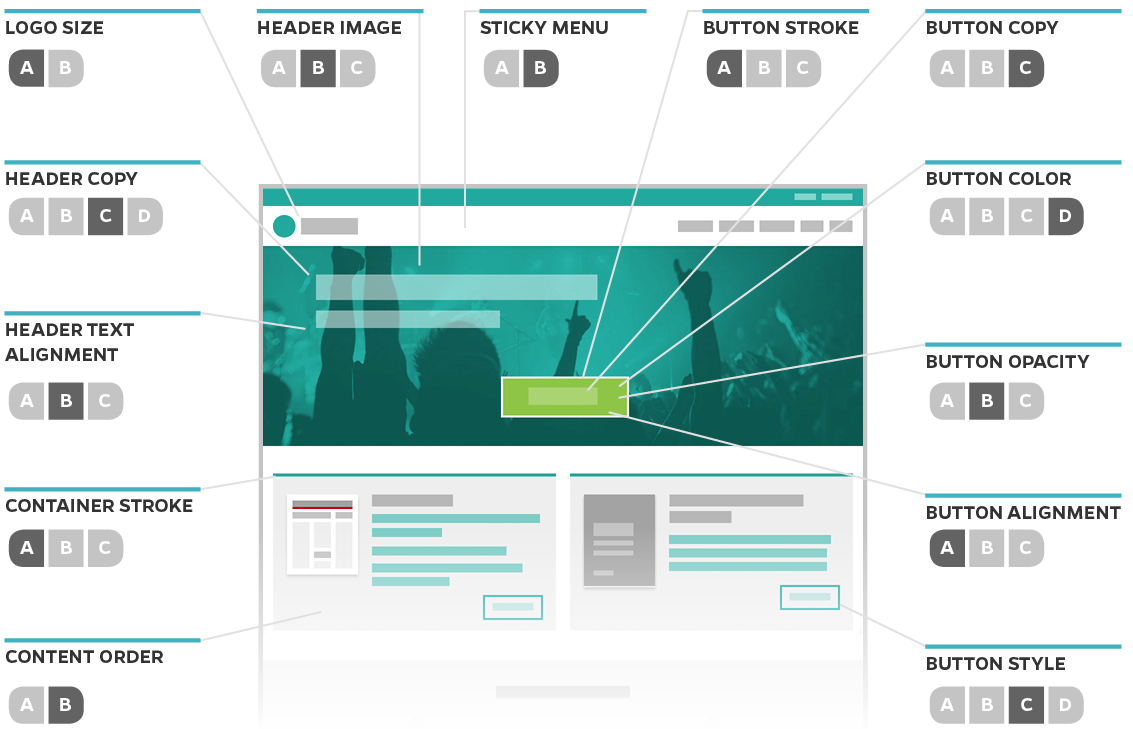} 
      \vspace{-0.6cm}
      \caption{Elements and Values of an Example Web Page Design. In
        this example, 13 elements each have 2-4 possible values,
        resulting in 1.1M combinations.
      }
      \label{fg:combinations}
    \end{center}
  \end{minipage}
  \hfill
  \begin{minipage}{\columnwidth}
    \begin{center}
      \includegraphics[width=\columnwidth]{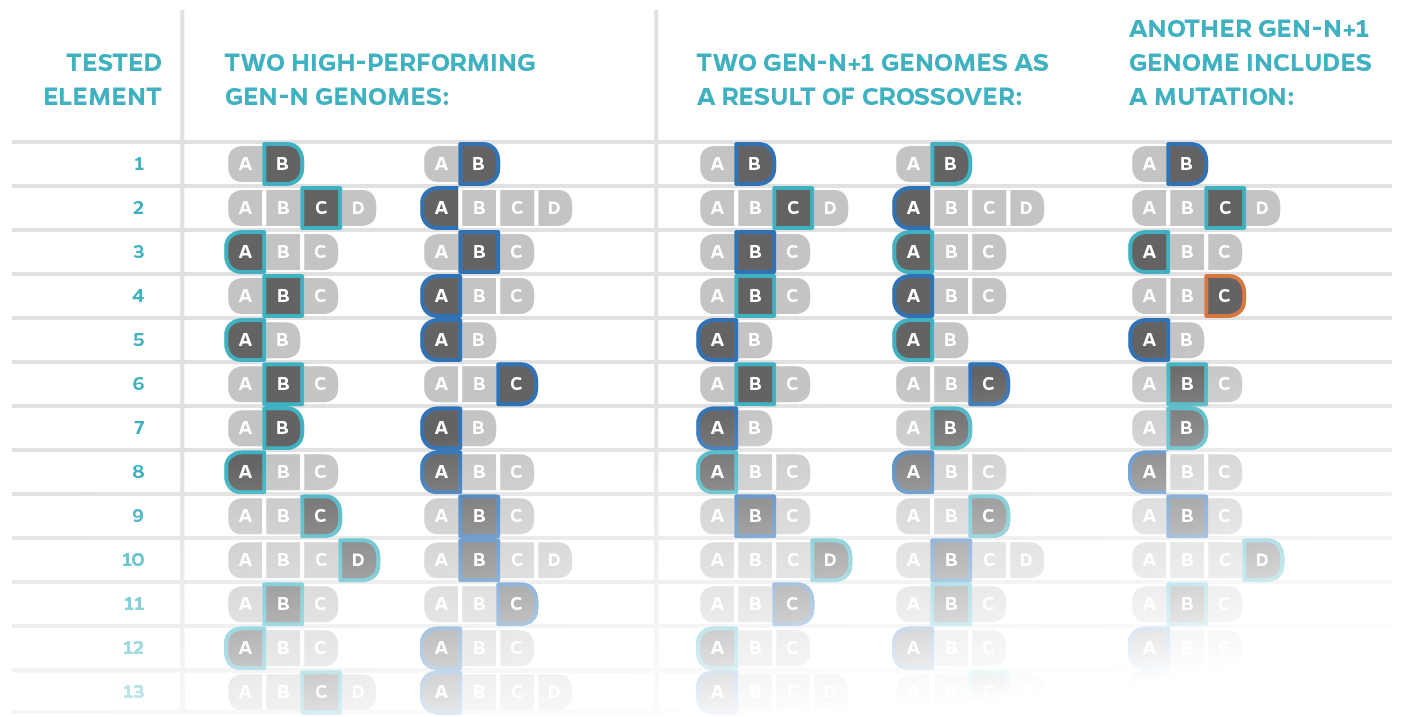}
      \vspace{-0.6cm}
      \caption{Genetic Encoding and Operations on Web Interface
        Candidates. The pages are represented as concatenations of their
        element values with one-hot encoding. Crossover and mutation
        operate on these vectors as usual, creating new combinations of
        values.
      }
      \label{fg:genomes}
    \end{center}
  \end{minipage}
\end{figure*}

\section{The Ascend Method}

Ascend consists of defining the space of possible web interfaces,
initializing the population with a good coverage of that space,
allocating traffic to candidates intelligently so that bad designs can
be eliminated early, and testing candidates online in parallel. Each
of these steps is described in more detail in this section.

\subsection{Defining the Search Space}

The starting point for Ascend is a search space defined by the web
designer. Ascend can be configured to optimize a design of a single
web-page, or a funnel consisting of multiple pages such as the landing
page, selections, and a shopping cart.  For each such space, the
designer specifies the elements on that page and values that they can
take.  For instance in the landing page example of
Figure~\ref{fg:combinations}, logo size, header image, button color,
content order are such elements, and they can each take on 2-4 values.

Ascend searches for good designs in the space of possible combinations
of these values. This space is combinatorial, and can be very large,
e.g. 1.1M in this example. Interestingly, it is exactly this
combinatorial nature that makes web-page optimization a good
application for evolution: Even though human designers have insight
into what values to use, their combinations are difficult to predict,
and need to be discovered by search process such as evolution.

\subsection{Initializing Evolution}
\label{sc:initialization}

A typical setup is that there is already a current design for the web
interface, and the goal for Ascend is to improve over its
performance. That is, the current design of the web interface is
designated as the Control, and improvement is measured compared to
that particular design.

Because fitness is evaluated with real users, exploration incurs real
cost to the customer. It is therefore important that the candidates
perform reasonably well throughout evolution, and especially in the
beginning.

If the initial population is generated randomly, many web interfaces
would perform poorly. Instead, the initial population is created using
the Control as a starting point: The candidates are created by
changing the value of one element systematically. In a small search
space, the initial population thus consists of all candidates with one
difference from the control; in a large search space, the population
is a sample of the set of such candidates. With such an
initialization, most of the candidates perform similarly to the
control. The candidates also cover the search dimensions well, thus
forming a good starting point for evolution.

\begin{figure*}[!t]
  \begin{center}
    \includegraphics[width=\textwidth]{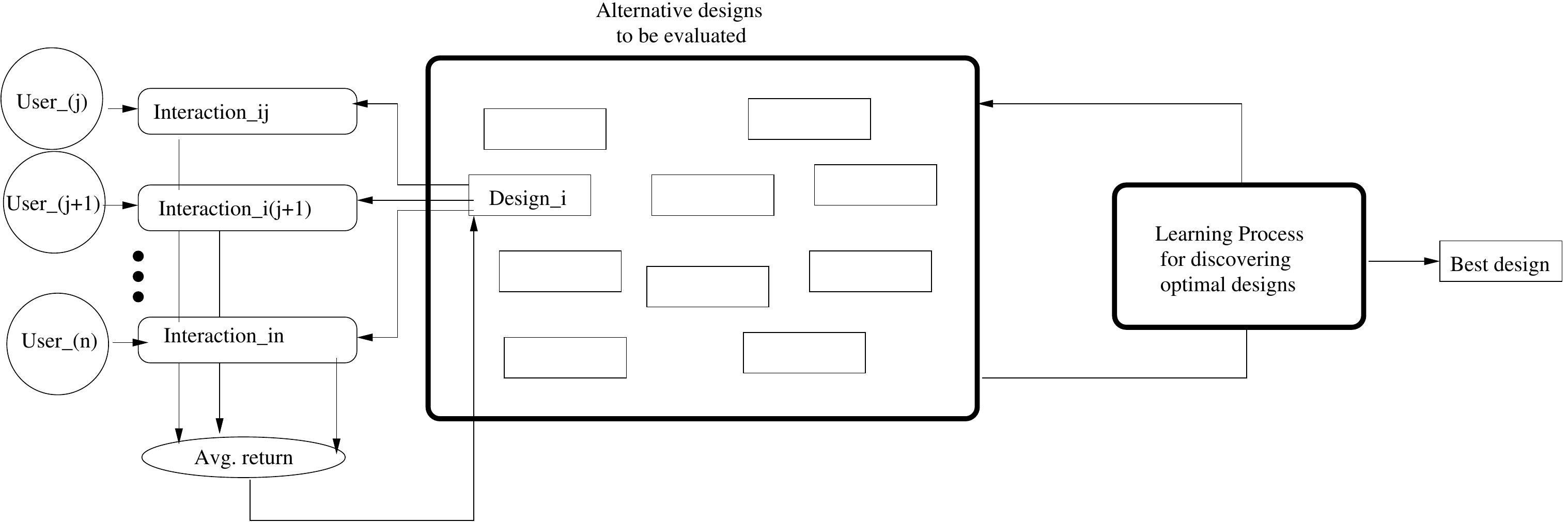} 
    \vspace{-0.6cm}
    \caption{Overall Architecture of the Online Evolution System. The
      outcome of each interaction (i.e.\ whether the user converted or
      not) constitutes one evaluation of a design. Many such
      evaluations $ij$ are run in parallel with different users $j$
      and averaged to estimate how good the design $i$ is. After all
      designs have been evaluated, the adaptation process discards bad
      designs and generates more variations of the best designs. This
      process of generation, testing, and selection is repeated until
      a sufficiently good design has been found or the time allocated
      for the process has been spent. The best design found so far is
      output as the result of the learning process.  The system thus
      discovers good designs for web interfaces through live online
      testing.  }
    \label{fg:designblockdiagram}
  \end{center}
\end{figure*}

\subsection{Evolutionary Process}
\label{sc:process}

Each page is represented as a genome, as shown for two example pages
in Figure~\ref{fg:genomes} (left side).  The usual genetic operations
such as crossover (re-combination of the elements in the two genomes;
middle) and mutation (randomly changing one element in the offspring;
right side) are then performed to create new candidates. In the
current implementation, fitness-proportionate selection is used to
generate offspring candidates from the current population. From the
current population of $n$ candidates, another $n$ new candidates are
generated in this way.

Because evaluations are expensive, consuming traffic for which most
customers have to pay, it is useful to minimize them during
evolution. Each page needs to be tested only to the extent that it is
possible to decide whether it is promising, i.e. whether it should
serve as a parent in the next generation, or should be discarded.  A
process similar to age-layering \cite{shahrzad:gecco16,hodjat:gptp13}
is therefore used to allocate fitness evaluations. At each generation,
each new candidate and each old candidate is evaluated with a small
number (a maturity age) of user interactions, such as 2000.  The top
$n$ candidates are retained, and the bottom $n$ discarded. In this
manner, bad candidates are eliminated quickly. Good candidates receive
progressively more evaluations, and the confidence in their fitness
estimate increases.

\begin{figure*}[!t]
  \begin{center}
    \includegraphics[width=\textwidth]{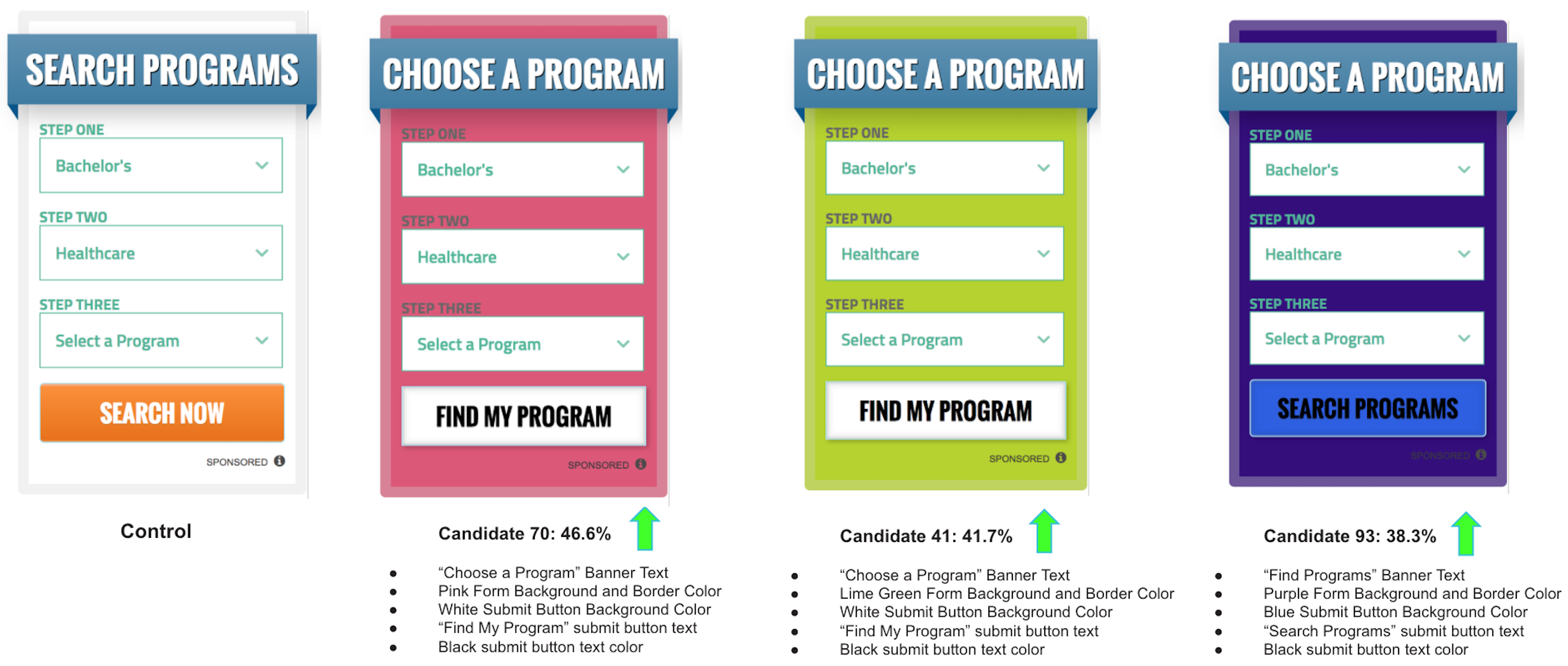} 
    \caption{The control design and three best evolved designs. After
      60 days of evolution with 599,008 user interactions, a design
      for the search widget was found that converted 46.6\% better
      than the control (5.61\% vs. 8.22\%), as well as other good
      designs. Much of the improvement was based on discovering a
      combination of colors that draws attention to the widget and
      makes the call to action clear.}
    \label{fg:abuvpages}
  \end{center}
\end{figure*}

In this process, Ascend learns which combinations of elements are
effective, and gradually focuses the search around the most promising
designs. It is thus sufficient to test only a tiny fraction of the
search space to find the best ones, i.e. thousands of pages instead of
millions or billions.

\subsection{Online Evolution}

While in simple cases (where the space of possible designs is small)
such optimization can potentially be carried out by simpler mechanisms
such as systematic search, hill-climbing, or reinforcement learning,
the population-based approach is particularly effective because the
evaluations can be done in parallel. The entire population can be
tested at once, as different users interact with the site
simultaneously.  It is also unnecessary to test each design to
statistical significance; only weak statistical evidence is sufficient
to proceed in the search. In this process, thousands of page designs
can be tested in a short time, which is impossible through A/B or
multivariate testing.

Figure~\ref{fg:designblockdiagram} shows the overall architecture of
the system. A population of alternative designs (center) are adapted
(right) based on evaluations with actual users (left).  The population
of designs (center) are evaluated with many users in parallel
(left). The evolutionary process (right) generates new designs and
outputs the best design in the end. The system also keeps track of
which design has been show to which user, so that they get to see the
same design if they return within a certain time limit (e.g. the same
day).

\section{Case Study}
\label{sec:casestudy}

\begin{figure*}[!t]
  \begin{center}
    \includegraphics[width=0.3\textwidth]{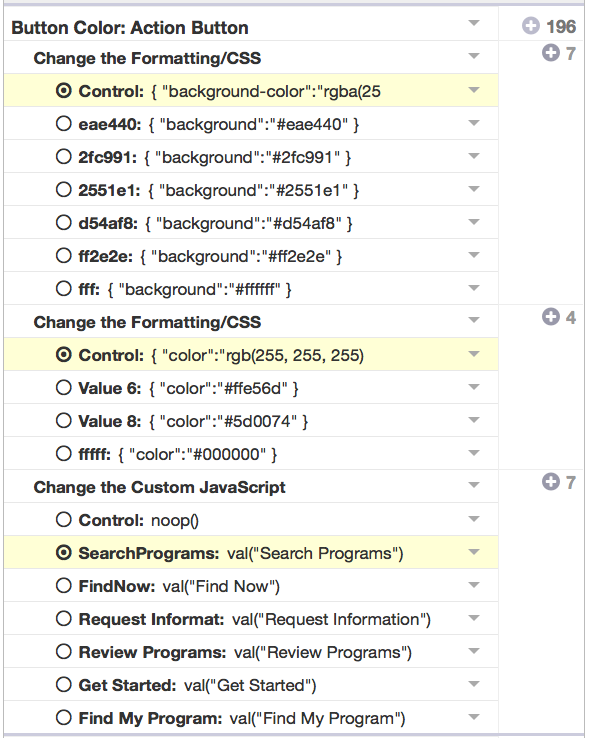} 
    \hfill
    \includegraphics[width=0.3\textwidth]{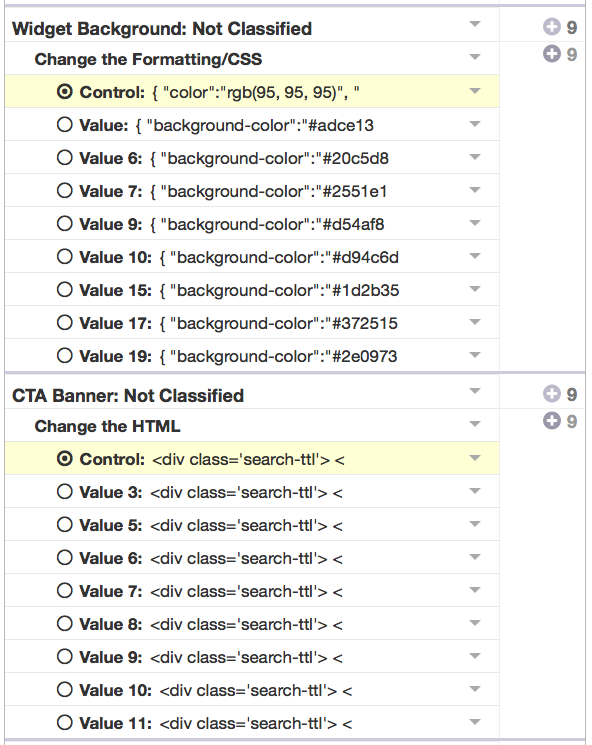} 
    \hfill
    \includegraphics[width=0.325\textwidth]{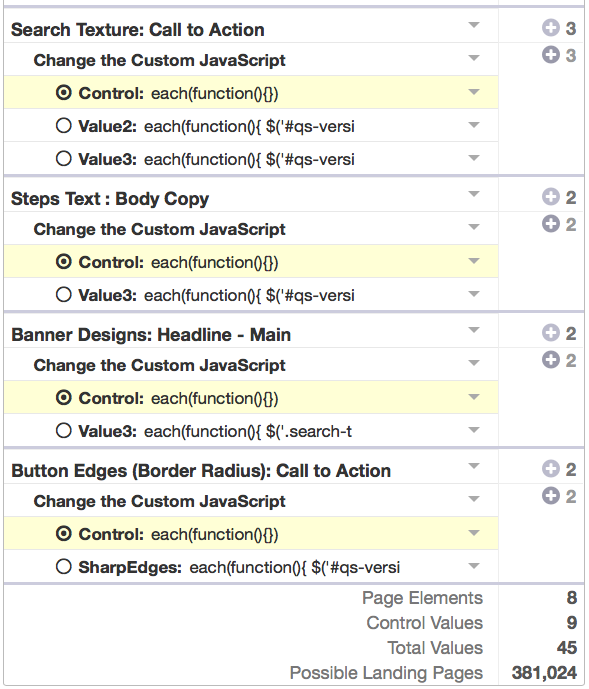} 
    \hfill
    \caption{A screenshot of the user interface for designing Ascend
      experiments, showing the elements and values in the
      learnhowtobecome.com case study. Nine elements with two to nine
      different values each result in 381,024 potential web page
      designs; the first value in each element is designated as the
      control.  This is a mid-size problem typical of current web
      interface designs.  }
    \label{fg:abuvspace}
  \end{center}
\end{figure*}

\begin{figure*}[!t]
  \begin{center}
    \includegraphics[width=\textwidth]{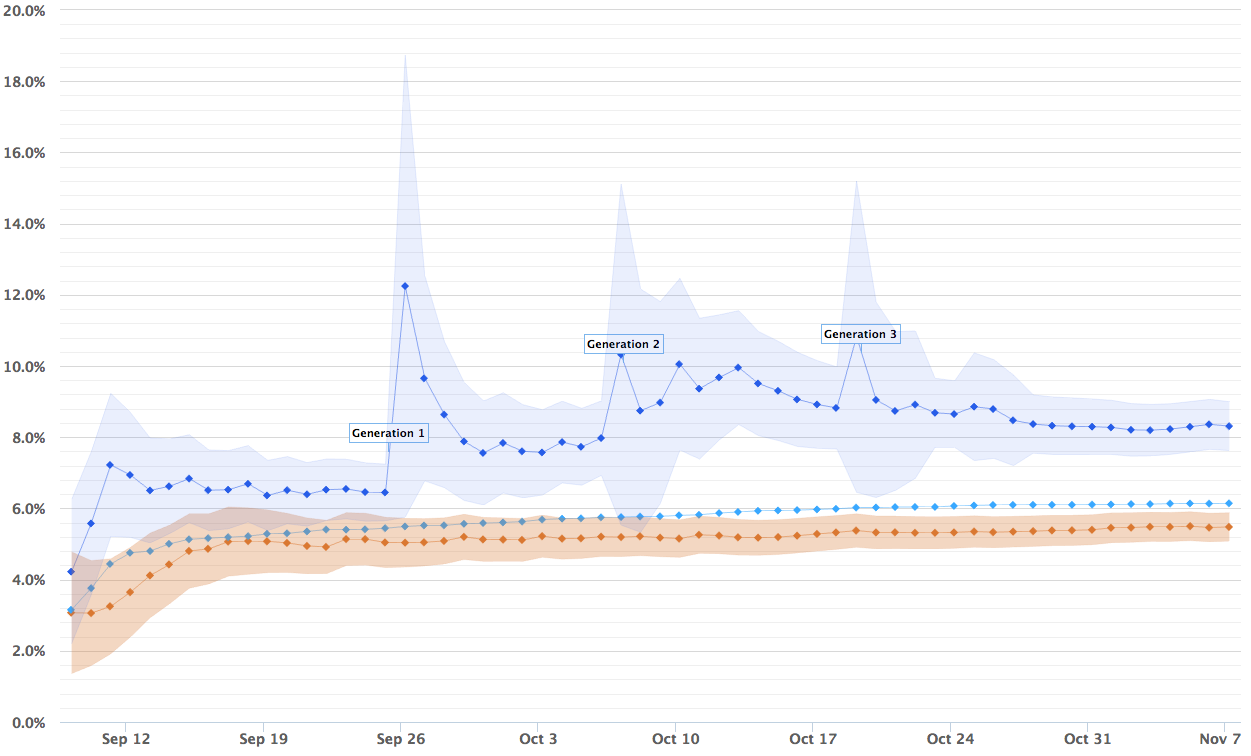} 
    \caption{Estimated Conversion Rates through the 60-day Online
      Evolution Run. Days are in the $x$-axis and the conversion rate
      on the $y$ axis. The dark blue dots (on top) indicate the
      current best candidate, the light blue dots (in the middle) an
      average of all currently active candidates, and the orange dots
      (at the bottom) the estimated performance of the control
      design. The shaded areas display the 95\% confidence intervals
      (from the binomial distribution with the observed mean). The
      dark blue peaks indicate the start of each new generation. Such
      peaks emerge because during the first few days, the new
      candidates have been evaluated only a small number of times, and
      some of them have very high estimated rates through random
      chance.  Eventually they will be evaluated in a maturity age of
      2000 user interactions, and the estimates become lower and the
      confidence intervals narrower. The elite candidates are tested
      across several generations as described in
      section~\ref{sc:process}, resulting in very narrow intervals
      towards the end.  Estimated conversion rates of the best
      candidates in later generations are significantly higher than
      control, suggesting that evolution is effective in discovering
      better candidates. Interestingly, the active population average
      is also higher than control, indicating that the experiment did
      not incur any cost in performance.}.
    \label{fg:abuvgenerations}
  \end{center}
\end{figure*}

As an example of how Ascend works, let us consider a case study on
optimizing the web interface for a media site that connects
users to online education programs.
%
%
This experiment was run in September through
November 2016 on the desktop traffic of the site.

The initial design for this page is shown in the left side of
Figure~\ref{fg:abuvpages}. It had been hand designed using standard
tools such as Optimizely. Its conversion rate during the time of the
experiment was found to be 5.61\%, which is typical of such web
interfaces. Based on this page, the web designers came up with nine
elements, with two to nine values each, resulting in 381,024 potential
combinations (Figure~\ref{fg:abuvspace}). While much larger search
spaces are possible, this example represents a mid-size space common
with many current sites.

\begin{figure*}[!t]
  \begin{center}
    \includegraphics[width=\textwidth]{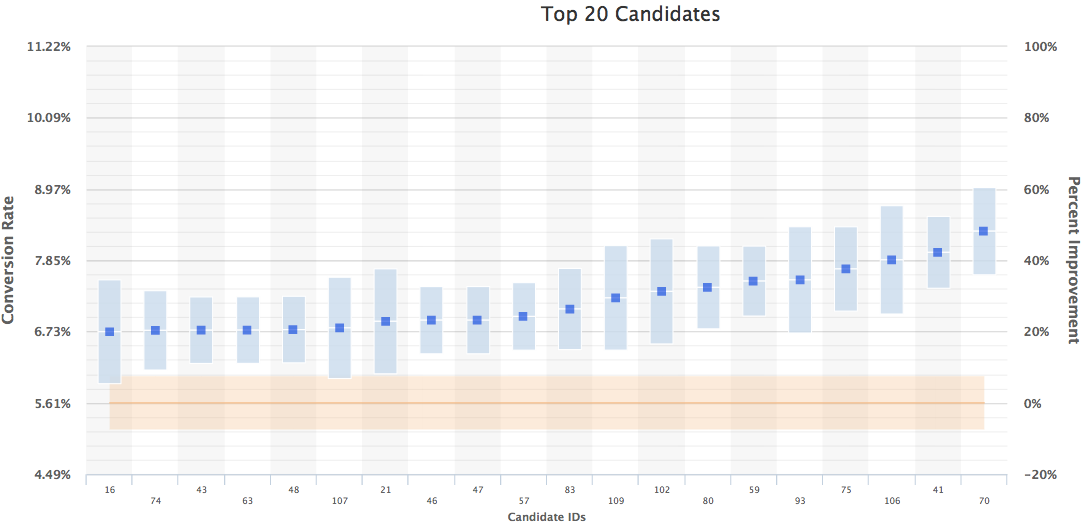} 
    \vspace{-0.6cm}
    \caption{Estimated Conversion Rates of the Top 20 Candidates
      vs.\ Control. The candidates are identified in the $x$-axis with
      numbers corresponding to those in
      Figure~\ref{fg:abuvpages}. They are ordered according to their
      estimated conversion rate. All these candidates are
      significantly better than the control at the 95\% level. The
      best one (on the right) is 46.6\% better, as confirmed with an
      independent A/B test that showed 43.5\% improvement.  }
    \label{fg:abuvtop}
  \end{center}
\end{figure*}

The initial population of 37 candidates was formed by systematically
replacing each of the values in the control page with one of the
alternative values, as described in
section~\ref{sc:initialization}. Evolution was then run for 60 days,
or four generations, altogether testing 111 candidates with 599,008
user interactions total.  The estimated conversion rates of the
candidates over this time are shown in
Figure~\ref{fg:abuvgenerations}. The conversion rates of the top 20
candidates are shown in Figure~\ref{fg:abuvtop}. These figures show
that evolution was successful in discovering significantly better
candidates than control.

As an independent verification, the three top candidates in
Figure~\ref{fg:abuvpages} were then subjected to an A/B test using
Optimizely. In about 6500 user interactions, the best candidate was
confirmed to increase the conversion rate by 43.5\% with greater than
99\% significance (and the other two by 37.1\% and 28.2\%)---which is
an excellent result given that the control was a candidate that was
already hand-optimized using state-of-the art tools.

Unlike Control, the top candidates utilize bright background colors to
draw attention to the widget.  There is an important interaction
between the background and the blue banner (whose color was
fixed)---in the best two designs (in the middle) the background is
distinct from the banner but not competing with it. Moreover, given
the colored background, a white button with black text provided the
most clear call for action. It is difficult to recognize such
interactions ahead of time, yet evolution discovered them early on,
and many of the later candidates built on them.  Other factors such as
an active call to action (i.e.\ ``Get Started'' and ``Find my
Program'' rather than ``Request Info'') amplified it further. At the
time evolution was turned off, better designs were still being
discovered, suggesting that a more prolonged evolution and a larger
search space (e.g. including banner color and other choices) could
have improved the results further.

\section{Future Work}
\label{sec:future}

Ascend has been applied to numerous web interfaces, and it has
consistently improved conversion rates by 20-200\% compared to hand
designed controls \cite{sentient:ascend17}. The main limitation is
often the human element: web designers, who are used to A/B and
multivariate testing, often try to minimize the search space, i.e.\
the number of elements and values, as much as possible, thereby not
giving evolution much space to explore and discover powerful
solutions. Often the evolution discovers significant improvement in a
couple of generations, and the designers are eager to adopt them right
away, instead of letting evolution optimize the designs
fully. Population-based optimization requires different thinking; as
designers become more comfortable with it, we believe they will let
evolution take its course, reaching more refined results.

Currently Ascend delivers one best design, or a small number of good
ones, in the end as the result, again in keeping with the A/B testing
tradition.  In many cases there are seasonal variations and other
long-term changing trends, making the performance of good designs
gradually decay. It is possible to counter this problem by running the
optimization again every few months. However, a new paradigm of
``always-on'' would be more appropriate: Evolutionary optimization can
be run continuously at a low volume, keeping up with changing trends
(i.e. through dynamic evolutionary optimization;
\cite{branke:book02}).  New designs can then be adopted periodically
when their performance exceeds old designs significantly.

Furthermore, currently Ascend optimizes a single design to be used
with all future users of a mobile or desktop site. An interesting
extension would be to take user segmentation \cite{yankelovich:hbr06}
into account, and evolve different pages for different kinds of
users. Moreover, such a mapping from user characterizations to page
designs can be automatic: A mapping system such as a neural network
can take user variables such as location, time, device, any past
history with the site as inputs, and generate the vector of elements
and their values as outputs. Neuroevolution
\cite{lehman:scholarpedia13,floreano:evolint08} can discover optimal
such mappings, in effect evolve to discover a dynamic, continuous
segmentation of the user space. Users will be shown designs that are
likely to convert well based on experience with other users with
similar characteristics, continuously and automatically. It will be
possible to analyze such evolved neural networks and discover what
variables are most predictive, characterize the main user segments,
and thereby develop an in-depth understanding of the opportunity.

Finally, the Ascend approach is not limited to optimizing conversions.
Any outcome that can be measured, such as revenue or user retention,
can be optimized. The approach can also be used in a different role,
such as optimizing the amount of resources spent on attracting users,
such as ad placement and selection, adword bidding, and email
marketing. The approach can be seen as a fundamental step in bringing
machine optimization into e-commerce, and demonstrating the value of
evolutionary computation in real-world problems.

\section{Conclusion}
\label{sec:conclusion}

Sentient Ascend demonstrates how interactive evolution can be scaled
up by testing a large number of candidates in parallel on real
users. It includes technology for keeping the cost of exploration
reasonable, and for minimizing the number of evaluations needed. From
the application point of view, Ascend is the first automated system
for massively multivariate conversion optimization---replacing A/B
with AI.  It finds the subtle combinations of variables that lead to
conversion increases.  The web designer can spend more time trying
ideas and less time doing statistics, giving them the freedom they
need to make a difference.


\bibliographystyle{ACM-Reference-Format}
\bibliography{/u/nn/bibs/nnstrings,/u/nn/bibs/nn,/u/risto/risto,newbib}


\begin{thebibliography}{00}


\ifx \showCODEN    \undefined \def \showCODEN     #1{\unskip}     \fi
\ifx \showDOI      \undefined \def \showDOI       #1{{\tt DOI:}\penalty0{#1}\ }
  \fi
\ifx \showISBNx    \undefined \def \showISBNx     #1{\unskip}     \fi
\ifx \showISBNxiii \undefined \def \showISBNxiii  #1{\unskip}     \fi
\ifx \showISSN     \undefined \def \showISSN      #1{\unskip}     \fi
\ifx \showLCCN     \undefined \def \showLCCN      #1{\unskip}     \fi
\ifx \shownote     \undefined \def \shownote      #1{#1}          \fi
\ifx \showarticletitle \undefined \def \showarticletitle #1{#1}   \fi
\ifx \showURL      \undefined \def \showURL       #1{#1}          \fi
\providecommand\bibfield[2]{#2}
\providecommand\bibinfo[2]{#2}
\providecommand\natexlab[1]{#1}
\providecommand\showeprint[2][]{arXiv:#2}

\bibitem[\protect\citeauthoryear{Ash, Page, and Ginty}{Ash
  et~al\mbox{.}}{2012}]%
        {ash:book12}
\bibfield{author}{\bibinfo{person}{Tim Ash}, \bibinfo{person}{Rich Page}, {and}
  \bibinfo{person}{Maura Ginty}.} \bibinfo{year}{2012}\natexlab{}.
\newblock \bibinfo{booktitle}{{\em Landing Page Optimization: The Definitie
  Guide to Testing and Tuning for Conversions\/} (\bibinfo{edition}{second}
  ed.)}.
\newblock \bibinfo{publisher}{Wiley}, \bibinfo{address}{Hoboken, NJ}.
\newblock


\bibitem[\protect\citeauthoryear{Brabham}{Brabham}{2013}]%
        {brabham:book13}
\bibfield{author}{\bibinfo{person}{Daren~C. Brabham}.}
  \bibinfo{year}{2013}\natexlab{}.
\newblock \bibinfo{booktitle}{{\em Crowdsourcing}}.
\newblock \bibinfo{publisher}{MIT Press}, \bibinfo{address}{Cambridge, MA}.
\newblock


\bibitem[\protect\citeauthoryear{Branke}{Branke}{2002}]%
        {branke:book02}
\bibfield{author}{\bibinfo{person}{J{\"u}rgen Branke}.}
  \bibinfo{year}{2002}\natexlab{}.
\newblock \bibinfo{booktitle}{{\em Evolutionary Optimization in Dynamic
  Environments}}.
\newblock \bibinfo{publisher}{Springer}, \bibinfo{address}{Berlin}.
\newblock


\bibitem[\protect\citeauthoryear{Builtwith}{Builtwith}{2017}]%
        {builtwith:jan17}
\bibfield{author}{\bibinfo{person}{Builtwith}.}
  \bibinfo{year}{2017}\natexlab{}.
\newblock \bibinfo{title}{{A/B} Testing Usage}.  (\bibinfo{year}{2017}).
\newblock
\showURL{%
\url{https://trends.builtwith.com/analytics/a-b-testing}}
\newblock
\shownote{Retrieved 1/9/2017.}


\bibitem[\protect\citeauthoryear{eMarketer}{eMarketer}{2016}]%
        {emarketer:sep16}
\bibfield{author}{\bibinfo{person}{eMarketer}.}
  \bibinfo{year}{2016}\natexlab{}.
\newblock \bibinfo{title}{US Digital Ad Spending to Surpass TV this Year}.
  (\bibinfo{year}{2016}).
\newblock
\showURL{%
\url{https://www.emarketer.com/Article/US-Digital-Ad-Spending-Surpass-TV-this-Year/1014469}}
\newblock
\shownote{Retrieved 2/1/2017.}


\bibitem[\protect\citeauthoryear{Floreano, D{\"u}rr, and Mattiussi}{Floreano
  et~al\mbox{.}}{2008}]%
        {floreano:evolint08}
\bibfield{author}{\bibinfo{person}{Dario Floreano}, \bibinfo{person}{Peter
  D{\"u}rr}, {and} \bibinfo{person}{Claudio Mattiussi}.}
  \bibinfo{year}{2008}\natexlab{}.
\newblock \showarticletitle{Neuroevolution: {F}rom Architectures to Learning}.
\newblock \bibinfo{journal}{{\em Evolutionary Intelligence\/}}
  \bibinfo{volume}{1} (\bibinfo{year}{2008}), \bibinfo{pages}{47--62}.
\newblock


\bibitem[\protect\citeauthoryear{Hodjat and Shahrzad}{Hodjat and
  Shahrzad}{2013}]%
        {hodjat:gptp13}
\bibfield{author}{\bibinfo{person}{Babak Hodjat} {and} \bibinfo{person}{Hormoz
  Shahrzad}.} \bibinfo{year}{2013}\natexlab{}.
\newblock \showarticletitle{Introducing an Age-Varying Fitness Estimation
  Function}.
\newblock In \bibinfo{booktitle}{{\em Genetic Programming Theory and Practice
  X}}, \bibfield{editor}{\bibinfo{person}{Rick Riolo},
  \bibinfo{person}{Ekaterina Vladislavleva}, \bibinfo{person}{Marylyn~D
  Ritchie}, {and} \bibinfo{person}{Jason~H. Moore}} (Eds.).
  \bibinfo{publisher}{Springer}, \bibinfo{address}{New York},
  \bibinfo{pages}{59--71}.
\newblock


\bibitem[\protect\citeauthoryear{Kohavi and Longbotham}{Kohavi and
  Longbotham}{2016}]%
        {kohavi:encyclopedia16}
\bibfield{author}{\bibinfo{person}{Ron Kohavi} {and} \bibinfo{person}{Roger
  Longbotham}.} \bibinfo{year}{2016}\natexlab{}.
\newblock \showarticletitle{Online Controlled Experiments and {A/B} Tests}.
\newblock In \bibinfo{booktitle}{{\em Encyclopedia of Machine Learning and Data
  Mining}}, \bibfield{editor}{\bibinfo{person}{Claude Sammut} {and}
  \bibinfo{person}{Geoffrey~I. Webb}} (Eds.). \bibinfo{publisher}{Springer},
  \bibinfo{address}{New York}.
\newblock


\bibitem[\protect\citeauthoryear{Lehman and Miikkulainen}{Lehman and
  Miikkulainen}{2013a}]%
        {lehman:tpnc13}
\bibfield{author}{\bibinfo{person}{Joel Lehman} {and} \bibinfo{person}{Risto
  Miikkulainen}.} \bibinfo{year}{2013}\natexlab{a}.
\newblock \showarticletitle{Boosting Interactive Evolution using Human
  Computation Markets}. In \bibinfo{booktitle}{{\em Proceedings of the 2nd
  International Conference on the Theory and Practice of Natural Computation}}.
  \bibinfo{publisher}{Springer}, \bibinfo{address}{Berlin}.
\newblock


\bibitem[\protect\citeauthoryear{Lehman and Miikkulainen}{Lehman and
  Miikkulainen}{2013b}]%
        {lehman:scholarpedia13}
\bibfield{author}{\bibinfo{person}{Joel Lehman} {and} \bibinfo{person}{Risto
  Miikkulainen}.} \bibinfo{year}{2013}\natexlab{b}.
\newblock \showarticletitle{Neuroevolution}.
\newblock \bibinfo{journal}{{\em Scholarpedia\/}} \bibinfo{volume}{8},
  \bibinfo{number}{6} (\bibinfo{year}{2013}), \bibinfo{pages}{30977}.
\newblock
\showURL{%
\url{http://nn.cs.utexas.edu/?lehman:scholarpedia13}}


\bibitem[\protect\citeauthoryear{Salehd and Shukairy}{Salehd and
  Shukairy}{2011}]%
        {salehd:book11}
\bibfield{author}{\bibinfo{person}{Khalid Salehd} {and} \bibinfo{person}{Ayat
  Shukairy}.} \bibinfo{year}{2011}\natexlab{}.
\newblock \bibinfo{booktitle}{{\em Conversion Optimization: {T}he Art and
  Science of Converting Prospects to Customers}}.
\newblock \bibinfo{publisher}{O'Reilly Media, Inc.},
  \bibinfo{address}{Sebastopol, CA}.
\newblock


\bibitem[\protect\citeauthoryear{Secretan, Beato, D'Ambrosio, Rodriguez,
  Campbell, Folsom-Kovarik, and Stanley}{Secretan et~al\mbox{.}}{2011}]%
        {secretan:ec11}
\bibfield{author}{\bibinfo{person}{Jimmy Secretan}, \bibinfo{person}{Nicholas
  Beato}, \bibinfo{person}{David~B. D'Ambrosio}, \bibinfo{person}{Adelein
  Rodriguez}, \bibinfo{person}{Adam Campbell}, \bibinfo{person}{J.~T.
  Folsom-Kovarik}, {and} \bibinfo{person}{Kenneth~O. Stanley}.}
  \bibinfo{year}{2011}\natexlab{}.
\newblock \showarticletitle{{P}icbreeder: {A} Case Study in Collaborative
  Evolutionary Exploration of Design Space}.
\newblock \bibinfo{journal}{{\em Evolutionary Computation\/}}
  \bibinfo{volume}{19} (\bibinfo{year}{2011}), \bibinfo{pages}{345--371}.
\newblock


\bibitem[\protect\citeauthoryear{{Sentient Technologies}}{{Sentient
  Technologies}}{2017}]%
        {sentient:ascend17}
\bibfield{author}{\bibinfo{person}{{Sentient Technologies}}}
  \bibinfo{year}{2017}\natexlab{}.
\newblock \bibinfo{title}{It's not {A/B}, I's {AI}}.  (\bibinfo{year}{2017}).
\newblock
\showURL{%
\url{http://www.sentient.ai/ascend}}
\newblock
\shownote{Retrieved 1/9/2017.}


\bibitem[\protect\citeauthoryear{Shahrzad, Hodjat, and Miikkulainen}{Shahrzad
  et~al\mbox{.}}{2016}]%
        {shahrzad:gecco16}
\bibfield{author}{\bibinfo{person}{Hormoz Shahrzad}, \bibinfo{person}{Babak
  Hodjat}, {and} \bibinfo{person}{Risto Miikkulainen}.}
  \bibinfo{year}{2016}\natexlab{}.
\newblock \showarticletitle{Estimating the Advantage of Age-Layering in
  Evolutionary Algorithms}. In \bibinfo{booktitle}{{\em Proceedings of the
  Genetic and Evolutionary Computation Conference (GECCO 2016)}}.
  \bibinfo{publisher}{ACM}, \bibinfo{address}{New York, NY, USA}.
\newblock


\bibitem[\protect\citeauthoryear{Takagi}{Takagi}{2001}]%
        {takagi:ieee01}
\bibfield{author}{\bibinfo{person}{H. Takagi}.}
  \bibinfo{year}{2001}\natexlab{}.
\newblock \showarticletitle{Interactive Evolutionary Computation: {F}usion of
  the Capacities of {EC} Optimization and Human Evaluation}.
\newblock \bibinfo{journal}{{\it Proc. IEEE}} \bibinfo{volume}{89},
  \bibinfo{number}{9} (\bibinfo{year}{2001}), \bibinfo{pages}{1275--1296}.
\newblock
\showURL{%
\url{http://ieeexplore.ieee.org/iel5/5/20546/00949485.pdf?tp=&arnumber=949485&isnumber=20546}}


\bibitem[\protect\citeauthoryear{Yankelovich and Meer}{Yankelovich and
  Meer}{2006}]%
        {yankelovich:hbr06}
\bibfield{author}{\bibinfo{person}{Daniel Yankelovich} {and}
  \bibinfo{person}{David Meer}.} \bibinfo{year}{2006}\natexlab{}.
\newblock \showarticletitle{Rediscovering Market Segmentation}.
\newblock \bibinfo{journal}{{\em Harvard Business Review\/}}
  \bibinfo{volume}{84}, \bibinfo{number}{2} (\bibinfo{year}{2006}).
\newblock


\end{thebibliography}

\end{document}